\documentclass[conference]{IEEEtran}
\usepackage{hyperref}
\usepackage{graphicx,psfrag,amsmath}
\usepackage{array,amssymb,amsfonts,epstopdf,mathtools,cite}
\usepackage{epsfig}
\usepackage{algpseudocode}
\newtheorem{proposition}{Proposition}
\newtheorem{lemma}{Lemma}
\newtheorem{observation}{Observation}
\newtheorem{corollary}{Corollary}

\newtheorem{theorem}{Theorem}
\newtheorem{example}{Example}

\def\mathbi#1{\textbf{\em #1}}
\newcommand{\goodchi}{\protect\raisebox{1.5pt}{$\chi$}}
\title{Error-Correcting Functional Index Codes, Generalized Exclusive Laws and Graph Coloring}

\begin{document}

\author{
\authorblockN{Anindya Gupta and B. Sundar Rajan}
\authorblockA{Department of Electrical Communication Engineering, Indian Institute of Science, Bangalore 560012, India\\ Email:\{anindya.g, bsrajan\}@ece.iisc.ernet.in}
}

\maketitle
\thispagestyle{empty}	
\begin{abstract}
We consider the \emph{functional index coding problem} over an error-free broadcast network in which a source generates a set of messages and there are multiple receivers, each holding a set of functions of source messages in its cache, called the \emph{Has-set}, and demands to know another set of functions of messages, called the \emph{Want-set}. Cognizant of the receivers' \emph{Has-sets}, the source aims to satisfy the demands of each receiver by making coded transmissions, called a \emph{functional index code}. The objective is to minimize the number of such transmissions required. The restriction a receiver's demands pose on the code is represented via a constraint called the \emph{generalized exclusive law} and obtain a code using the \emph{confusion graph} constructed using these constraints. Bounds on the size of an optimal code based on the parameters of the confusion graph are presented. Next, we consider the case of erroneous transmissions and provide a necessary and sufficient condition that an FIC must satisfy for correct decoding of desired functions at each receiver and obtain a lower bound on the length of an error-correcting FIC.
\end{abstract}
\section{Introduction}

There has been an increasing interest in the index coding problem (ICP) because of its potential to afford throughput gain in ad hoc wireless networks. It finds commercial application in dissemination of popular multimedia content as in IPTV, DVB, P2P file sharing. An instance of ICP, $\mathcal{I(X,R)}$, comprises a single source/transmitter possessing a set of messages, $X=\{ x_1,x_2,\ldots,x_K\}$, and a set of clients/receivers, $R=\{ R_1,R_2,\ldots,R_N\}$. Each client, $R_i=(H_i,W_i)$, knows a subset of messages, $H_i \subset X$, a priori, and demands to know another subset of messages, $W_i \subset X$, where, $H_i \cap W_i = \emptyset$. These two sets are respectively named the \emph{Has-set} and the \emph{Want-set} of the client. The transmitter can broadcast functions of messages in $X$ to the clients via a noiseless channel. The objective is to equip the transmitter with the minimum number of encoding functions such that the demands of all the clients is satisfied upon reception of the same. Such a situation may arise, for example, when a satellite or a broadcasting station wishes to transmit a large file (message set) to many receivers by breaking it into multiple fragments (messages). Some receivers may miss out certain messages due to multitude of reasons including bad or intermittent signal reception, interference from other sources, channel noise, power outage, temporary equipment failure, and bad weather. Instead of retransmitting missed out messages, the transmitter can take cognizance of what the receivers already have in their cache and transmit fewer coded messages so that the demand of each receiver is satisfied.

Alternatively, the ICP can be posed as a problem of source coding with side information available at receivers and objective is to design a code of minimum size. For example, consider the ICP depicted in Table\,\ref{table1}, where $x_k \in \mathbb{F}_2$, $\forall k \in \{1,2,\ldots,K\}$.
\begin{table}[h]
\centering
\setlength\extrarowheight{1pt}
\begin{tabular}{|c|c|c|}
\hline
\textbf{Client} & \textbf{Has-set} & \textbf{Want-set} \\
\hline
\textbf{$R_1$} & $\{x_5\;x_2\}$ & $\{x_1\}$\\
\hline
\textbf{$R_2$} & $\{x_1\;x_3\}$ & $\{x_2\}$\\
\hline 
\textbf{$R_3$} & $\{x_2\;x_4\}$ & $\{x_3\}$\\
\hline
\textbf{$R_4$} & $\{x_3\;x_5\}$ & $\{x_4\}$\\
\hline
\textbf{$R_5$} & $\{x_4\;x_1\}$ & $\{x_5\}$\\
\hline 
\end{tabular}
\caption{}
\label{table1}
\vspace{-20pt}
\end{table}

It can be verified that three transmissions, viz., $x_1+x_2$, $x_3+x_4$ and $x_5$ would suffice (all operations over $\mathbb{F}_2$). When messages are elements of $\mathbb{F}_2^2$ (each message is a 2-bit word), \emph{i.e.}, $x_k=(x_k^1,x_k^2)$, $x_k^1,x_k^2 \in \mathbb{F}_2$, then the following set of transmission also suffices: $\{ x_1^1+x_2^1,\: x_2^2+x_3^1,\: x_3^2+x_4^1,\:x_4^2+x_5^1,\:x_5^2+x_1^2 \}$. This economizes the number of transmissions, by saving one bit, when compared to the former scheme which requires transmitting six bits. These two schemes are example of what are called scalar and vector linear index codes (since the encoding operations are linear), respectively.

\subsection{Related Work and Motivation}
The functional source coding with side-information problem (FSCSIP), wherein the receiver wishes to compute a function, $f(X,Y)$, of its side information random variable, $Y$, and the source random variable, $X$, was studied in \cite{OrlRoch} using the characteristic graph of the problem instance. An optimal vertex coloring of the characteristic graph obtained from the problem instance was shown to provide a minimum size code in \cite{Med}. The extension of this problem to multiple receiver case was subsequently dealt in \cite{MedMulti}, wherein each receiver possessed multiple random variables correlated to source as side information and demanded several functions of source and their side information. In \cite{Anin}, we proposed and studied a variant of the FSCSIP wherein the receiver demands and holds as side information functions of source messages. 

The ICP was introduced in \cite{BK} and a method to obtain index code based on partial clique cover of the underlying side information graph was proposed, which was  further studied in \cite{BBJK} using graph theory. The main conjecture of \cite{BBJK} that linear index codes are always optimal was refuted in \cite{LubStav}. Advantages of block/vector coding were established in \cite{ICRel,BCSI}. In \cite{BCSI}, it was shown that a minimum size index code can be obtained from a vertex coloring of confusion graph of the ICP. Finding a minimum size index codes is NP-hard \cite{BBJK,MinTx}. Several heuristic solutions were provided in \cite{MinTx,CIC,EffSoln}. Error-correcting index codes were introduced and studied in \cite{Dau}. The case where side information includes linear combination of messages was first studied in \cite{BCCSI}, which was motivated by the fact that some clients may still fail to receive some coded transmissions due to reasons mentioned earlier and transmitter may need to compute a new index code after every transmissions taking into account the updated caches and demands of the receivers. Error-correcting index codes for this case were proposed in \cite{Marco}. This motivated us to study problems with arbitrary functions as side information.

Network coding problem has garnered much attention of the research community, see \cite{Yeung} and references therein. The main advantage network coding offers is improvement in throughput by exploiting the fact that intermediate nodes can perform computation on incoming information rather than merely route them. Though ICP falls as a special case of a more general network coding problem, equivalence between the two has been shown in \cite{ICRel,Eqv}, \emph{i.e.}, every index coding problem can be converted to an instance of network coding problem and vice versa. Network coding capacity by examining the corresponding index coding problem was studied in \cite{Eqv}. The in-network function computation problem comprises source nodes generating messages, intermediate nodes performing computation on incoming information and sink nodes seeking functions of source messages \cite{Dey,Appu}. The aim is to maximize the frequency of target function computation per network use \cite{Appu}. This motivated to study ICP where clients' demands may also include functions of messages. 
 
\subsection{Contributions and Organization}
The contributions and organization of the paper are as follows:
\begin{enumerate}
\item In Section~III, we propose and study the \emph{functional index coding with side information problem} (FICP) wherein there is one transmitter which generates a finite number of messages, each taking value form a finite field and there are multiple receivers, each knowing a set of functions of source messages and demanding a different set of functions of source messages. The objective is to transmit a \emph{functional index code} (FIC) over a broadcast channel of minimum length so that demands of each receiver is met. The notions of the \emph{generalized exclusive law} (GEL), which a functional index code must satisfy, and \emph{confusion graph} are defined. The FICP generalizes the following two problems:
\begin{enumerate}
\item The conventional ICP: The clients know and demand subset of messages.
\item The FSCSIP of \cite{Anin}: There is only one client which knows and demands functions of source messages.
\end{enumerate}
In \cite{Anin}, a code for an FSCSIP was obtained using the associated \emph{row-Latin rectangle} (RLR). An RLR is a table with \emph{Has-values} indexing the rows and the \emph{Want-values} indexing the columns and a message vectors appear in a cell if it evaluates to the row and column index of that cell. Two message vectors in the same row but different columns should be mapped to different codewords \cite[Theorem 2]{Anin}. For multiple-users, there will be multiple RLRs and the above constraint must be simultaneously satisfied for each of them. We attempted to obtain FICs using the RLR approach with no success. So, we use graph theoretic approach in this paper to obtain FICs.
\item In Section~IV, we show that a FIC must satisfy the GELs of each receiver so that their demands can be met. We obtain such a code by coloring the vertices of the confusion graph of the FICP. For single-receiver case, \emph{i.e.}, FSCSIP, satisfying GEL (Proposition 1) is shown to be same as satisfying \cite[Theorem 2]{Anin} and so, vertex coloring approach can also be used to obtain codes for FSCSIP. Some properties of the confusion graph are given in Section~IV-A and bounds on the optimal code size are obtained using these properties. Some illustration of the proposed technique are given in Section~IV-B.
\item In Section~V, we consider transmission over a channel that introduces at most $\delta$ errors and provide a necessary and sufficient condition that an FIC must satisfy so that the receivers can correctly obtain the values of functions in its \emph{Want-set}.  We also provide the \emph{Singleton bound} for error-correcting FIC (linear or non-linear). Some examples of optimal error-correcting FICs (both satisfying and not satisfying) the Singleton bound are given.
\end{enumerate}
Relevant concepts from graph theory are introduced in Section~II and the paper is concluded with a discussion on scope of further work in Section~VI.

\section{Preliminaries}
In this section, we present some concepts from graph theory relevant to our work. The reader is referred to \cite{Grph1,Grph2,Grph3} and references therein for further details.

A graph is a pair $\mathcal{G=(V,E)}$, where $\mathcal{V}$ is the set of vertices/nodes and $\mathcal{E}$ $\subseteq\mathcal{V}$ $\times \mathcal{V}$ is the set of edges. A graph is said to be undirected if the edges have no orientation, \emph{i.e.}, edges $(v_1,v_2)$ and $(v_2,v_1)$ are indistinguishable. A simple graph is an undirected graph without loops (edges originating and terminating at the same node) and without multiple edges between nodes. Set of neighboring vertices of a vertex $v$ is denoted by $N(v)$. 
An independent set is a subset of vertices such that no two vertices in the subset are adjacent. The size of a largest independent set is called the independence number and denoted by $\alpha(G)$. A component of graph is a subgraph in which there is a path between any two vertices and none of its vertices are connected to vertices not in this subgraph. 
A complete multipartite graph is one whose vertex set can be partitioned into several subsets such that there is no edge between vertices from the same partition class and an edge between them if they are from different classes. A regular graph is one in which each vertex has the same number of neighbors.

A vertex coloring of a graph $\mathcal{G}$ is an assignment of colors to its vertices such that no two adjacent vertices are like-colored, \emph{i.e.}, it is a surjective map $c:\mathcal{V}\longrightarrow C$, where $C$ is called the set of colors, such that $c(v_i)\neq c(v_j)$ if $(v_i,v_j)\in \mathcal{E}$. Such a coloring is called a $|C|$-coloring of $\mathcal{G}$. A vertex coloring stratifies vertices of a graph into disjoint subsets called color classes such that no two adjacent vertices are in the same class. The minimum number of colors required to color a graph is called its chromatic number, and is denoted by $\goodchi(\mathcal{G})$. A $\goodchi(\mathcal{G})$-coloring is referred to as a minimum vertex coloring of $\mathcal{G}$. Finding $\goodchi(\mathcal{G})$ of a general graph is an NP-hard problem. The fractional chromatic number is the minimum ratio $p/q$ such that there exists $p$ independent sets $V_1,V_2,\ldots,V_p$ (not necessarily distinct), with each vertex contained in exactly $q$ of them. Equivalently, given colors with some weight fractions, fractional coloring is assignment of a subset of color to vertices such that adjacent vertices have no color in common and sum of weight fraction of colors assigned to each vertex is at least one. The sum of weight fractions of the fractional coloring that uses fewest colors is the fractional chromatic number. Four colorings of the 5-cycle graph are given below; even though some colorings use more colors, the fractional chromatic number remains same or decreases. Its chromatic number and fractional chromatic numbers are $3$ and $5/2$ respectively. Applications of vertex coloring include solving scheduling problems, computer register allocation, bandwidth allocation to various users, finding channel codes of specified minimum distance and solving sudokus.  
\begin{figure}[htbp]
\centering
\includegraphics[scale=0.65]{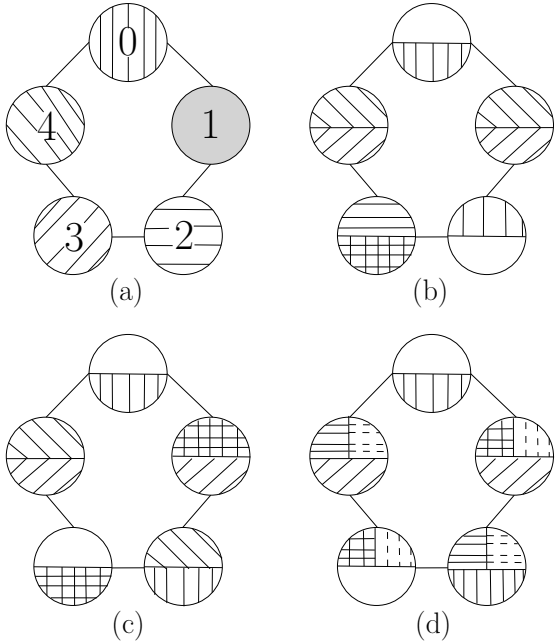}
\caption{(a) An optimal coloring with $3$ colors, (b) A suboptimal $6/2$ fractional coloring, (c) An optimal $5/2$ fractional coloring with $5$ colors and (d) An optimal $5/2$ coloring with $7$ colors $(3\cdot 1/2 + 4\cdot 1/4=5/2)$.}
\label{fig_cycle}
\end{figure}

The graph sum of two graphs, $G_1=(V,E_1)$ and $G_2=(V,E_2)$, on the same set of vertices is the graph $G_1+G_2=(V,E_1\cup E_2)$. 
The graph disjoint union of two graphs $(V_1,E_1)$ and $(V_2,E_2)$ with disjoint vertex sets is $(V_1\cup V_2,E_1\cup E_2)$; there will be no edges between elements of $V_1$ and $V_2$. The OR or co-normal product $G^2$ of $G=(V,E)$ with itself has $V^2$ as the vertex set and two distinct vertices $(u_1,u_2)$ and $(v_1,v_2)$ are adjacent iff $u_1$ and $v_1$ are adjacent or $u_2$ and $v_2$ are adjacent in $G$ or both. Similarly, the vertex set of $G^n$ is $V^n$ and two distinct vertices $(v_1,v_2,\ldots,v_n)$ and $(u_1,u_2,\ldots,u_n),\, v_i,u_i\in V,\forall i\in [n],$ are adjacent if $v_i$ and $u_i$ are adjacent in $G$ for at least one $i\in [n]$.

An automorphism of a graph $(V,E)$ is a permutation $f$ of its vertices, such that a pair of vertices $(a,b)$ form an edge iff the pair $(f(a),f(b))$ also form an edge. A graph, $(V,E)$, is said to be symmetric if, given any two edges $(a,b)$ and $(x,y)$, there exists an automorphism, $f:V\rightarrow V$, such that $f(a)=x$ and $f(b)=y$, \emph{i.e.}, every pair of adjacent vertices can be mapped into any other pair of adjacent vertices by an automorphism. A graph is vertex-transitive if every vertex can be mapped to any other vertex by an automorphism. Every symmetric graph is vertex-transitive and every vertex-transitive graph is regular.

Let $(G,\circ)$ be a finite abelian group with identity $e$. Let $S$ be a subset of $G$ such that $e\notin S$ and if $s$ is in $S$ then so is its inverse. The Cayley graph of $G$ with the \textit{connection set} $S$ is a graph with elements of $G$ as the vertices and each vertex $g$ is connected to $|S|$ other vertices, \emph{viz.}, $\{g\circ s:\,s\in S\}$. Thus, a Calyey graph is an undirected simple (since $e\notin S$) regular (each vertex has $|S|$ neighbors) graph. Every Cayley graph is vertex-transitive. For example, the $5$-cycle graph of Fig.~\ref{fig_cycle}(a) is the Cayley graph of $\mathbb{Z}_5$ with the connection set $\{1,4\}$.

For a simple undirected graph $G=(V,E)$, following are some properties of the graph parameters discussed above that we will use in Section~IV-A \cite{BCSI,Grph2,Grph3,Grph4}
\begin{align}
\alpha (G^n) &\;=\; (\alpha (G))^n\\
\goodchi _f(G)&\;\leqslant \; \goodchi (G) \leqslant \goodchi _f(G)(1+\log \alpha (G))\\
\goodchi _f(G)&\; \geqslant \; \textstyle\frac{|V(G)|}{\alpha (G)}\\
\goodchi _f(G^n) &\;=\; (\goodchi _f(G))^n
\end{align}
For vertex-transitive graphs, equality in (3) holds.

\begin{figure*}[t]
\begin{align}
\label{el_examp}
\begin{aligned}
\mathcal{M}(x_1,x_2,x_3)\neq \mathcal{M}(x_1',& x_2',x_3'),\: \mathrm{if} \: x_1=x_1' \: \mathrm{and}\: (x_2+x_3,x_1+x_3)\neq (x_2'+x_3',x_1'+x_3')\\
\mathcal{M}(x_1,x_2,x_3)\neq \mathcal{M}(x_1',x_2',x_3'),&\: \mathrm{if}\: \mathit{Maj}(x_1,x_2,x_3)=\mathit{Maj}(x_1,x_2,x_3),\: \mathrm{and}\: (x_1,x_2,x_3)\neq (x_1',x_2',x_3')
\end{aligned}
\end{align}
\hrule
\vspace{-10pt}
\end{figure*}
\section{Network Model}
In this section, we formally define the functional index coding problem, where the clients are permitted to hold as side information and/or demand functions of source messages rather than knowing a priori and demanding copies of messages only as in the conventional ICP. 

Throughout the paper it is assumed that source generates $K$ (finite) messages and there are $N$ client nodes. The set $\{1,2,\ldots,r\}$ is denoted by $[r],\;r\in \mathbb{N}$. A message, $x_k, k\in [K]$, is assumed to be an $n$-tuple over a finite $q$-ary field, $\mathbb{F}_q^n$, \emph{i.e.}, $x_k=(x_{k,1},x_{k,2},\ldots,x_{k,n})$, where $x_{k,j}$ is the $j^{th}$ sub-packet of the $k^{th}$ message and $x_{k,j}\in \mathbb{F}_q,\;\forall j\in [n]$ and some $n\in \mathbb{N}$. A vector $x=(x_1,x_2,\ldots,x_K)\in \mathbb{F}_q^{nK}$ is considered as an $nK$-tuple over $\mathbb{F}_q$ and is referred to as a message vector. We use $h_{i,j}$ and $w_{i,l}$ to denote functions in the \textit{Has-set} and \textit{Want-set} of the $i^{th}$ client, respectively, where $h_{i,j},w_{i,l}:\mathbb{F}_q^{nK} \longrightarrow \mathbb{F}_q^{n},\;\forall \; i\in [N],\;j,l\in \mathbb{N}$. We refer to functions in the \emph{Has-} \emph{(Want-) set} as the \emph{Has (Want) functions}. Union of disjoint subsets is denoted using $\sqcup$. Entropy of a random variable $X$ is denoted by $\mathbi{H}(X)$. The problem considered in this paper is defined below.

\textit{Definition 1 (Functional Index Coding Problem):} An instance of FICP, $\mathcal{F(X,R)}$, consists of:\\
1. A transmitter with a message set $X = \{x_1,x_2,\ldots,x_K\}$, where $\;x_k\in \mathbb{F}_q^n,\forall k\in [K]$.\\ 
2. A set of clients/receivers, $R=\{ R_1,R_2,\ldots,R_N\}$, where, $\forall i\in [N]$, $R_i=(H_i,W_i),\;H_i=\{ h_{i,1},h_{i,2},\ldots ,h_{i,|H_i|}\}$ and $W_i=\{w_{i,1},w_{i,2},\ldots ,w_{i,|W_i|}\}$ where $h_{i,j},w_{i,l}:\mathbb{F}_q^{nK} \longrightarrow \mathbb{F}_q^{n},\;\forall \; i\in [N],\;j,l\in \mathbb{N}$..

The FICP where $h_{\ast,\ast}\,s$ and $w_{\ast,\ast}\,s$ are equal to some message in $X$ correspond to the conventional ICP. The FICP where $h_{\ast,\ast}\,s$ are linear combinations of messages and $w_{\ast,\ast}\,s$ are equal to some message in $X$ was considered in \cite{BCCSI}. Thus, the above definition subsumes the ICP studied so far as special cases. Define $H_i(x)\triangleq(h_{i,1}(x),h_{i,2}(x),\ldots ,h_{i,|H_i|}(x))$ and $W_i(x)\triangleq(w_{i,1}(x),W_{i,2}(x),\ldots ,W_{i,|W_i|}(x))$, where $x=(x_1,x_2,\ldots,x_K)\in \mathbb{F}_q^{nK}$ as the \emph{Has-} and \emph{Want-value} for $x$. Let $\mathcal{H}_i$ and $\mathcal{W}_i$ be the set of all possible \emph{Has-} and \emph{Want-values} of the $i^{th}$ receiver. When we write $H_i(x)=0$ $(W_i(x)=0)$, $0$ denotes the all-zero $n|H_i|$ $(n|W_i|)$ length vector. When all the \textit{Has-} (\textit{Want-}) \textit{functions} of a client (say $R_i$) are linear, we represent them using a matrix $M_{H_i}\in \mathbb{F}_q^{nK\times n|H_i|}$ $(M_{W_i}\in \mathbb{F}_q^{nK\times n|W_i|})$ wherein the $j^{th}$ column contains the coding coefficients of the $j^{th}$ \textit{Has-} (\textit{Want-}) \textit{function} and $H_i(x)=xM_{H_i}$ $(W_i(x)=xM_{W_i})$. When all the \textit{Has-} and \textit{Want-functions} at all the receivers are linear, we call it a linear FICP.

\begin{example}
\label{ex_ICP}
Consider the FICP given in Table\,\ref{table_ICP}.
\begin{table}[htbp]
\setlength\extrarowheight{1pt}
\centering
\begin{tabular}{|c|c|c|}
\hline
\textbf{Client} & \textbf{Has-set} & \textbf{Want-set} \\
\hline
\textbf{$R_1$} & $\{x_1\}$ & $\{x_2+x_3,x_1+x_3\}$\\
\hline 
\textbf{$R_2$} & $\{\mathit{Maj}(x_1,x_2,x_3)\}$ & $\{x_1,x_2,x_3\}$\\
\hline 
\end{tabular}
\vspace{5pt}
\caption{}
\label{table_ICP}
\vspace{-10pt}
\end{table}
\noindent In this example, $q=2,n=1,K=3,N=2$, $\mathit{Maj}$ denotes the majority function, addition is over $\mathbb{F}_2$, $X=\{x_1,x_2,x_3\}$, $h_{1,1}=x_1,\, h_{2,1}=\mathit{Maj}(x_1,x_2,x_3),\, w_{1,1}=x_2+x_3,\, w_{1,2}=x_1+x_3$, and $w_{2,1}=x_1,\, w_{2,2}=x_2,\,  w_{2,3}=x_3$.\hfill $\square$
\end{example}
\textit{Definition 2 (Functional Index Code):} A functional index code (FIC) for a given $\mathcal{F(X,R)}$ comprises of:\\
1. An encoding map, $\mathcal{M}:\mathbb{F}_q^{nK}\longrightarrow \mathcal{B},\,\mathcal{B}\subseteq \mathbb{F}_q^L$, for some $L\in \mathbb{N}$\\
2. Decoding functions, $\mathcal{D}_i:\mathcal{B} \times \mathbb{F}_q^{n|H_i|}\longrightarrow \mathbb{F}_q^{n|W_i|}$, such that $\forall i\in [N]$ and $\forall x\in \mathbb{F}_q^{nK}$, $\mathcal{D}_i(\mathcal{M}(x),H_i(x))=W_i(x)$.
The set $\mathcal{B}$ is the codebook and $L=\lceil log_q|\mathcal{B}|\rceil$ is referred to as the length of the FIC. The transmitter broadcasts the $L$ length codewords and the receivers use their respective decoding maps to obtain the desired functions. A linear FIC can be represented using a matrix $M\in \mathbb{F}_q^{nK\times L}$; the $j^{th}$ column contains the coding coefficients of the $j^{th}$ coded transmission.

The length of the code need not be a multiple of $n$ emphasizing that vector coding is also considered. This was observed in the vector solution of ICP of Table\,\ref{table1} ($n=2,L=5$). The elements of the set $\mathcal{B}$ are referred to as codewords. The objective is to minimize $L$, or equivalently $\mathcal{B}$, to achieve maximum throughput gain. A code which achieves minimum possible $L$ is said to be \emph{optimal}. We denote the optimal length by $L_{opt}$. For the functional ICP, a code is said to be \emph{perfect} if $L=\mu(\mathcal{F})\triangleq max_{i\in [N]}\lceil \max _{h\in \mathcal{H}_i} \mathbi{H}(W_i|H_i=h)\rceil$ (cf. \cite{Anin}). For the conventional ICP, $\mathbi{H}(W_i|H_i=h)=\mathbi{H}(W_i)=|W_i|,\forall i\in [N],\forall h\in \mathcal{H}_i$, since \emph{Has-set} and \emph{Want-set} are disjoint sets of independent messages. Thus, the definition of perfect index code, given in \cite{ICRel}, for the conventional ICP falls as a special case of our definition. Arguments similar to those in \cite{ICRel} can be used to verify that $\mu(\mathcal{F})$ bounds the number of transmissions from below for a given FICP.

\begin{example}
\label{ex_IC}
Continuing with Example\,\ref{ex_ICP}, it can be verified that transmitting $(x_1+x_3,\, x_2+x_3)$ satisfies demands of both clients and this is a perfect FIC. \hfill $\square$
\end{example}

Depending upon the clients' side information and demands, the transmitter attempts to formulate an optimal FIC. Put differently, the transmitter chooses a many-to-one map $\mathcal{M}:\mathbb{F}_q^{nK}\longrightarrow \mathcal{B}\subseteq \mathbb{F}_q^L$. To meet every client's demands, the map should satisfy a set of constraints dictated by $\mathcal{F(X,R)}$. These constraints, that we refer to as the generalized exclusive laws, are defined below.

\textit{Definition 3 (Generalized Exclusive Laws):} For successful decoding of demands of the $i^{th}$ client, the index coding map should be such that $\forall x\neq x'\in \mathbb{F}_q^{nK}$, $\mathcal{M}(x)\neq \mathcal{M}(x')$ whenever $H_i(x)=H_i(x')$ and $W_i(x)\neq W_i(x')$. 

We refer to this constraint as the $i^{th}$ generalized exclusive law (GEL) for $\mathcal{F(X,R)}$ and denote it by $\mathit{E_i}(\mathcal{F})$. An FICP prescribes $N$ such GELs, one for each receiver, that the FIC must satisfy so that all clients can reconstruct desired information unambiguously. 

\begin{example}
\label{ex_EL}
For the FICP given in Example\,\ref{ex_ICP}, the GELs prescribed by $R_1$ and $R_2$ are given in (\ref{el_examp}) at the top of this page.
\hfill $\square$
\end{example}

The above definition can be viewed as a generalization of mutually exclusive laws used to obtain broadcast maps in a wireless bidirectional relaying scenario\cite{VvR1,SR,VvRCol}.

\textit{Definition 4 (Confusion Graph):} The confusion graph of an FICP $\mathcal{F(X,R)}$, denoted by $\mathcal{C(F)}$, is a simple undirected graph whose vertex set is $\mathcal{V}=\{ 0,1,\ldots,q^{nK}-1\}$, and the edge set is $\mathcal{E}=\{ (x,x')\in \mathcal{V}^2:H_i(x)=H_i(x')$ and $W_i(x)\neq W_i(x')$ for some $i \in [N]\}$.

The adjacent $nK$-tuples are said to be \textit{confusable}. Thus, the vertex set corresponds to $q^{nK}$ possible message vectors and the edge set corresponds to all possible pairs of message vectors that must be mapped to different codewords by the encoding map as required by the GELs. 

\begin{example}
\label{ex_CG}
We continue with Example \ref{ex_EL} and construct $\mathcal{C(F)}$ for FICP of Example \ref{ex_ICP}. The confusion graph is shown in Fig.\,\ref{fig_CG}. Decimal equivalents of 3-bit message vectors are used to label the vertices. \hfill $\square$

\begin{figure}[htbp]
\centering
\includegraphics[scale=0.92]{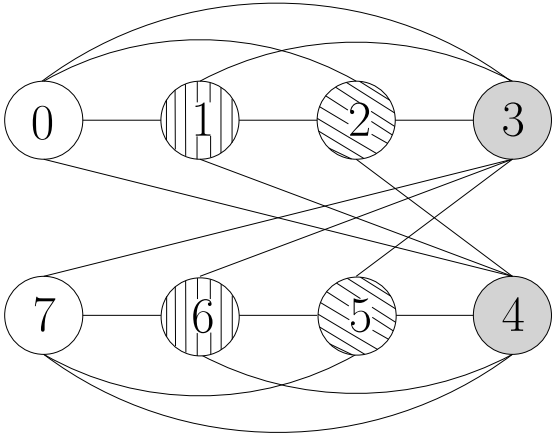}
\caption{Confusion graph of FICP of Table\,\ref{table_ICP}}
\label{fig_CG}
\end{figure}

\end{example}

The confusion graph constructed using the above definition will be identical to that of \cite{BCSI} for the case when side information and demands of the clients include only messages, \emph{i.e.}, the conventional ICP. Such a formulation obviates the construction of the directed hypergraph representation of ICP suggested in \cite{BCSI} or replacement of a receiver with $|W_i|>1$ with $|W_i|$ receivers each with singleton \emph{Want-sets} \cite{BCSI,ICRel}. Furthermore, none of the directed hypergraph \cite{BCSI}, side information graph \cite{BBJK,OngHo}, information flow graph \cite{OngHo} or bipartite graph \cite{BipIC} representation can be used to represent a FICP.

\section{Results and Illustrations}
In this section, we state and prove our results, provide an algorithm to construct an optimal FIC, specify some properties of the confusion graph and obtain bounds on the code size for a linear FICP.

\begin{proposition} The demands of the $i^{th}$ user can be met iff the $i^{th}$ GEL is satisfied by the FIC.\\
\textit{Proof:} Assume that the source generated a particular message vector $x\in \mathbb{F}_q^{nK}$. Let $x'\neq x \in \mathbb{F}_q^{nK}$ be such that $H_i(x)=H_i(x')$ and $W_i(x)\neq W_i(x')$.\\
\textit{a}) (Necessary part) If $\mathcal{M}(x)=\mathcal{M}(x')$, \emph{i.e.}, same codeword is assigned to both $x$ and $x'$, then there are two possible decoder outputs at the $i^{th}$ user, $W_i(x)$ and $W_i(x')$. Since both these possibilities are different, the FIC fails to satisfy the $i^{th}$ user.\\
\textit{b}) (Sufficient part) If $\mathcal{M}(x)\neq \mathcal{M}(x')$, \emph{i.e.}, different codewords are assigned to $x$ and $x'$, then, given $H_i(x)$, the $i^{th}$ user can uniquely identify $W_i(x$) when the source broadcasts $\mathcal{M}(x)$. Let $x''\neq x \in \mathbb{F}_q^{nK}$ be such that $H_i(x)\neq H_i(x'')$, $W_i(x)\neq W_i(x'')$ and $\mathcal{M}(x)=\mathcal{M}(x'')$. In this case, $H_i(x)$ assists in decoding to $W_i(x)$ and not to $W_i(x'')$. \hfill $\blacksquare$
\end{proposition}
Hence, it follows that in order to satisfy the demands of all the users, the FIC must satisfy all the GELs simultaneously.

For a single-user case, \emph{i.e.}, for an FSCSIP, if we construct a RLR using \cite[Definition 5]{Anin}, a pair of confusable message vectors will be in same row but different columns. Thus, a code for FSCSIP satisfying \cite[Definition 5]{Anin} will also satisfy Proposition 1 and vice versa.

\begin{proposition} For a given $\mathcal{F(X,R)}$, encoding maps that satisfy all the GELs simultaneously can be obtained from a vertex coloring of the confusion graph, $\mathcal{C(F)}$.\\
\textit{Proof:} Consider a vertex coloring of $\mathcal{C(F)}$ using $c$ colors. Since a vertex coloring outputs disjoint subsets of vertices such that no two adjacent vertices are in the same class, \emph{i.e.}, $\mathcal{V}=V_1 \sqcup V_2 \sqcup \ldots \sqcup V_c$, the vertices corresponding to confusable message vectors are colored using different colors. An FIC can be obtained by assigning one codeword to message vectors corresponding to vertices in the same color class. The size of the code thus found is $c$. \hfill $\blacksquare$
\end{proposition}

The size of an optimal FIC equals the chromatic number, $\goodchi(\mathcal{C})$, of the confusion graph $\mathcal{C(F)}$ and the length $L=\lceil log_q\goodchi(\mathcal{C})\rceil$. If $L>log_q\goodchi(\mathcal{C})$, then all possible $q^L$ codewords will not be required; different choices of $\goodchi(\mathcal{C})$ out of $q^L$ possibilities will lead to different optimal codes.

Method to construct the confusion graph and obtain a code for an FICP is given in Algorithm 1.
\begin{figure}[h]
\hrule height 1pt
\vspace*{2pt}
\begin{algorithmic}[1]
\Statex \textbf{Algorithm 1:} \textbf{FIC}($\mathcal{F,A},n$). Algorithm to construct $\mathcal{C(F)}$ and find an FIC
\vspace*{2pt}
\hrule
\vspace*{2pt}
\Statex \textbf{Input:} FICP $\mathcal{F(X,R)}$, $\mathbb{F}_q$, $n$ 
\Statex \textbf{Initialize:} $K=|\mathcal{X}|$, $N=|\mathcal{R}|$, $\mathcal{C(F)=(V,E)}$, $\mathcal{E}=\emptyset$, $\mathcal{V}=\{0,1,\ldots,q^{nK}-1\}$
\For{each $(x,x'),\: x\neq x'\in \mathbb{F}_q^{nK}$}
\For{each user $R_i,\: i\in [N]$}
\If{$H_i(x)=H_i(x')$ \textbf{and} $W_i(x)\neq W_i(x')$}
\State $\mathcal{E}=\mathcal{E}\cup (x,x')$
\State \textbf{break}
\EndIf
\EndFor
\EndFor
\State Color $\mathcal{C(F)}$ and obtain the color classes $(V_1,V_2,\ldots ,V_c)$ 
\State Set $L=\lceil \log_q{c} \rceil $ 
\State Choose $\mathcal{B}\subset\mathbb{F}_q^L,\:\mathcal{B}=\{y_1,y_2,\ldots,y_{c}\}$
\For{each $l\in [c]$}
\State $\mathcal{M}(x)=b_l,\:\forall x\in V_l$
\EndFor
\Statex \textbf{Output:} FIC $\mathcal{M}:\mathbb{F}_q^{nK} \longrightarrow \mathcal{B}$
\end{algorithmic}
\hrule height 1pt
\end{figure}

A brief description of Algorithm 1 is given below:
\begin{enumerate}
\item[1.] Initialize $\mathcal{C(F)}$ to be an edgeless graph on $q^{nK}$ nodes. Lines 1--8 add edges to $\mathcal{C(F)}$ iteratively as follows: for each unordered pair of distinct message vectors, if any of $N$ GELs forbid them to be mapped onto the same codeword, then add an edge between nodes corresponding to those message vectors. At most $2N$\,${q^{nK}}\choose{2}$ comparisons are to be made to obtain the confusion graph.
\item[2.] Color $\mathcal{C(F)}$ and obtain the color classes, viz., $V_1,V_2,\ldots ,V_c$. Since graph vertex coloring is, in general, an NP-hard problem, heuristics may be used to do the same; the resulting coloring and hence the code may not be optimal \cite{Grph1,Grph2}.
\item[3.] Assign vertices/message vectors in the same color class to a single codeword (Lines 12--14). The FIC thus obtained is optimal iff the coloring algorithm returns a minimum vertex coloring, \emph{i.e.}, $c=\goodchi(\mathcal{C})$.
\end{enumerate}

\begin{figure}[htbp]
\centering
\includegraphics[scale=0.8]{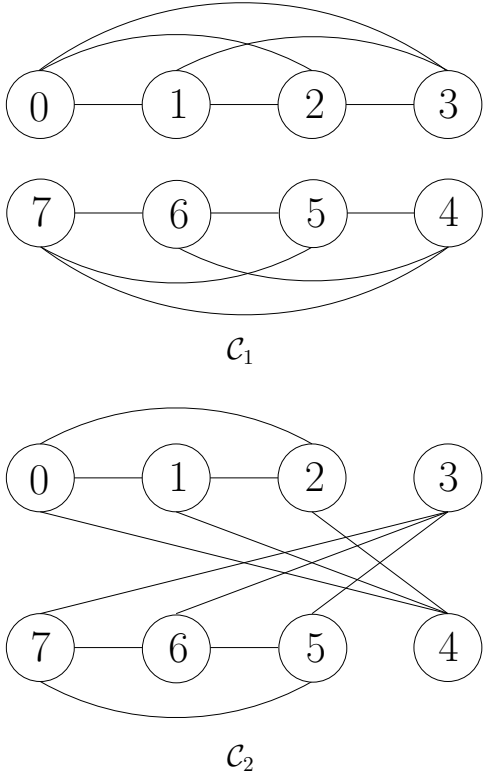}
\caption{Confusion graph of $R_1$ and $R_2$ of FICP of Table\,\ref{table_ICP}}
\label{fig_CG2}
\end{figure}

The confusion graphs of the two receivers and the vertex colored confusion graph of FICP of Table~\ref{table_ICP} are given in Fig.~\ref{fig_CG2} and Fig.~\ref{fig_CG} respectively; code size is $4$ ($=\goodchi(\mathcal{C})$) and code length is $2$. 
Two possible codeword assignments are given below.

\vspace*{-7pt}
\begin{small}
\begin{equation*}
\begin{aligned}
\{0,7\}\rightarrow 00 \enskip \{1,6\}\rightarrow 01 \enskip \{2,5\}\rightarrow 10 \enskip \{3,4\}\rightarrow 11 \\
\{0,7\}\rightarrow 01 \enskip \{1,6\}\rightarrow 10 \enskip \{2,5\}\rightarrow 00 \enskip \{3,4\}\rightarrow 11
\end{aligned}
\end{equation*}
\end{small}
The left and right assignments correspond to transmitting $(x_1+x_2,x_1+x_3)$ and $(x_1+x_3,1+x_2+x_3)$ respectively.

When messages take value from $\mathbb{F}_2^n,\,n\geqslant 1$, closed-form expressions for the transmissions as functions of messages can be obtained after codewords assignment (see \cite{Anin} and references therein).

\subsection{Properties of the Confusion Graph}
Some observations regarding the confusion graphs are given below. Upper and lower bounds on the size of code $(|\mathcal{B}|)$ is obtained using these properties.

\begin{observation} Let $\mathcal{C}_{i}$ denote the confusion graph of the $i^{th}$ receiver when the block length of each message is $1$. Then the confusion graph of the FICP for the scalar case is $\mathcal{C(F)}=\mathcal{C}_{1}+\mathcal{C}_{2}+\ldots+ \mathcal{C}_{N}$. When the message block length is $n\:(>1)$, the confusion graph of the $i^{th}$ receiver is $\mathcal{C}_{i}^n$ and that of the FICP is $\mathcal{C(F)}^n=(\mathcal{C}_{1}+\ldots+ \mathcal{C}_{N})^n= \mathcal{C}_{1}^n+\ldots+ \mathcal{C}_{N}^n$. The chromatic number of $\mathcal{C(F)}^n$ is then the optimal code size $|\mathcal{B}|$. For the $n$-fold OR product of a graph with itself \cite{Grph3},
\begin{equation*}
\lim _{n\rightarrow \infty} (\goodchi(G^n))^{1/n}=\goodchi _f(G).
\end{equation*}
Thus, from (2) we infer that increasing $n$ may lead to reduction of the code size $|\mathcal{B}|$.
 \hfill $\triangleleft$
\end{observation}
\textit{Remark:} Let $L_{(n)}$ denote the length of an FIC (not necessarily optimal) when block length is $n$. An FIC when message block length is $n$ can be obtained by splitting a message into several sub-blocks of smaller block lengths and encoding them separately; this technique may give suboptimal codes. In other words, if $n=n_1+n_2+n_3$, then an FIC for block length $n$ can be obtained by clustering the $n$ sub-packets of messages in groups of $n_1$, $n_2$ and $n_3$ and then encoding each cluster separately so that $L_{(n)}=L_{(n_1)}+L_{(n_2)}+L_{(n_3)}$\footnote{A partition of a non-negative integer $n$ is a representation of $n$ as a sum of other non-negative integers (ordering is irrelevant); \emph{e.g.}, there are $7$ ways of partitioning $5$, \emph{viz.}, $5=1+1+1+1+1=1+1+1+2=1+2+2=1+1+3=2+3=1+4$. The partition function $p(n)$ denotes the number of ways $n$ can be partitioned, \emph{e.g.}, $p(5)=7$.}. For example, for $n=5$, an FIC can be obtained by encoding each sub-packet separately in which case $L_{(5)}=5L_{(1)}$, or splitting it into sub-blocks of lengths $2$ and $3$ and encoding them separately, in which case $L_{(5)}=L_{(2)}+L_{(3)}$; in fact there are $7$ possible ways of doing this.

\begin{lemma} For a linear FICP, the confusion graph of each receiver will be a Cayley graph.\\
\textit{Proof:} Let $S=\mathsf{Null}(M_H)\cap \{\mathbb{F}_q^{nK}\backslash \mathsf{Null}(M_W)\}\subset \mathbb{F}_q^{nK}$, \textit{i.e.}, $S=\{s\in\mathbb{F}_q^{nK}: \,sM_H=0$ and $sM_W\neq 0\}$. Note that the additive identity $(\mathbf{0})$ of $\mathbb{F}_q^{nK}$ is not in $S$ (since $\mathbf{0}\cdot M_W=0$) and for every $s\in S$ its inverse is also in $S$ (since $sM_H=0$ and $sM_W\neq 0$ implies $-sM_H=0$ and $-sM_W\neq 0$ respectively). Consider the Caley graph of $(\mathbb{F}_q^{nK},+)$ with the connection set $S$; a vertex $x$ will be connected to $|S|$ vertices, \emph{viz.}, $N(x)=\{x+s:\, s\in S\}$. Note that $(x+s)M_H=xM_H$ and $(x+s)M_W\neq xM_W$. Hence, $N(x)$ is the set of message vectors confusable with $x$ and this Cayley graph is the confusion graph of the said receiver.
\hfill $\blacksquare$
\end{lemma}

\begin{lemma} For a linear FICP, the confusion graph will be a Cayley graph of $\mathbb{F}_q^{nK}$.\\
\textit{Reason:} Let $S_i$ be the connection set of the Cayley graph $\mathcal{C}_i$ of the $i_{th}$ receiver. Since $\mathcal{C(F)}=G_1+G_2+\ldots+G_N$, $\mathcal{C(F)}$ will also be a Cayley graph of $\mathbb{F}_q^{nK}$ with the connection set $\cup _{i\in [N]}S_i$. The graph will consequently be vertex-transitive and regular. Some bounds on the chromatic number of vertex-transitive and regular graphs are given in \cite{vt,reg}.
\hfill $\blacksquare$
\end{lemma}

\begin{example}
Consider the following linear FICP with $4$ messages over $\mathbb{F}_2$ with $2$ receivers.
\begin{table}[h]
\setlength\extrarowheight{1pt}
\centering
\begin{tabular}{|c|c|c|}
\hline
\textbf{Client} & \textbf{Has-set} & \textbf{Want-set} \\
\hline
\textbf{$R_1$} & $\{x_1+x_2\}$ & $\{x_2+x_3\}$\\
\hline 
\textbf{$R_2$} & $\{x_3+x_4\}$ & $\{x_1+x_4,\,x_1+x_2+x_3+x_4\}$\\
\hline 
\end{tabular}
\vspace{5pt}
\caption{}
\vspace{-20pt}
\end{table}
Connection sets for the two receivers are $S_1=\{0010,0011,1100,1101\}$ and $S_2=\{1100,$ $1000,0011,0111,0100,1011\}$ respectively, and that for the confusion graph of the FICP is $S=S_1\cup S_2$. \hfill $\square$
\end{example}

\begin{theorem}
For an FICP, the size of an optimal codebook is bounded as follows: 
\begin{equation}
\left(\goodchi _f(\mathcal{C})\right)^n \;\; \leqslant \;\; |\mathcal{B}| \;\; \leqslant \;\; \left(\goodchi _f(\mathcal{C})\right)^n (1+n\log \alpha (\mathcal{C}) )
\end{equation}
where $\mathcal{C}$ is the confusion graph of $\mathcal{F(X,R)}$ for the scalar case, $\alpha (\mathcal{C})$ is its independence number and messages are of block length $n$.\\
\textit{Proof:} Using (4) in (2) and the fact that $|\mathcal{B}|=\goodchi (\mathcal{C})$, we get the desired result. \hfill $\blacksquare$
\end{theorem}

\begin{corollary}
The size of an optimal codebook for a linear FICP $\mathcal{F(X,R)}$ is bounded as follows:
\begin{equation}
\left(\frac{q^K}{\alpha (\mathcal{C})}\right)^n \;\; \leqslant \;\; |\mathcal{B}| \;\; \leqslant \;\; \left(\frac{q^K}{\alpha (\mathcal{C})}\right)^n (1+n\log \alpha (\mathcal{C}) ).
\end{equation}
\\
\textit{Proof:} Since the confusion graph of a linear FICP is vertex-transitive, using (1) and (3) we have
\begin{align}
\goodchi _f (\mathcal{C}^n) \;\; = \;\; \frac{q^{nK}}{\alpha (\mathcal{C}^n)}\;\;=\;\;\frac{q^{nK}}{(\alpha (\mathcal{C}))^n}
\end{align}
Substituting (8) in (2) we get
\begin{equation*}
\left(\frac{q^K}{\alpha (\mathcal{C})}\right)^n \;\; \leqslant \;\; |\mathcal{B}| \;\; \leqslant \;\; \left(\frac{q^K}{\alpha (\mathcal{C})}\right)^n (1+n\log \alpha (\mathcal{C}) ).
\end{equation*}
This also gives bounds on the code size required for the variant of ICP studied in \cite{BCCSI}. \hfill $\blacksquare$
\end{corollary}

\noindent \textit{Remark:} Theorem~1 and Corollary~1 generalize Theorem~1.1 of \cite{BCSI} which bounds the code size of the conventional ICPs.


\begin{figure*}[t]
\begin{align}
\label{el_has}
\begin{aligned}
\mathcal{M}(x_1,x_2,x_3,x_4)&\neq \mathcal{M}(x_1,x_2',x_3',x_4'),\;\mathrm{if}\:(x_2,x_3,x_4)\neq (x_2,x_3',x_4')\\
\mathcal{M}(x_1,x_2,x_3,x_4)&\neq \mathcal{M}(x_1',x_2,x_3',x_4'),\;\mathrm{if}\:(x_1,x_3,x_4)\neq (x_1',x_3',x_4')\\
\mathcal{M}(x_1,x_2,x_3,x_4)&\neq \mathcal{M}(x_1',x_2',x_3,x_4'),\;\mathrm{if}\:(x_1,x_2,x_4)\neq (x_1',x_2',x_4')\\
\mathcal{M}(x_1,x_2,x_3,x_4)&\neq \mathcal{M}(x_1',x_2,x_3',x_4),\;\mathrm{if}\:(x_1,x_3,x_4)\neq (x_1',x_3',x_4')\\
\mathcal{M}(x_1,x_2,x_3,x_4)\neq \mathcal{M}(x_1',x_2',x_3',x_4'),&\;\mathrm{if}\:\mathit{Maj}(x_1,x_2,x_3)=\mathit{Maj}(x_1',x_2',x_3')\:\mathrm{and}\:(x_1,x_2,x_3,x_4)\neq (x_1',x_2',x_3',x_4')
\end{aligned}
\end{align}
\hrule
\vspace{-12pt}
\end{figure*}

\begin{figure*}[t]
\begin{align}
\label{el_want}
\begin{aligned}
\mathcal{M}(x_1,x_2,x_3,x_4,x_5)&\neq \mathcal{M}(x_1,x_2,x_3,x_4',x_5'),\;\mathrm{if}\:(x_4,x_5)\neq (x_4',x_5')\\
\mathcal{M}(x_1,x_2,x_3,x_4,x_5&)\neq \mathcal{M}(x_1',x_2',x_3',x_4,x_5),\;\mathrm{if}\:W_2(x)\neq W_2(x)
\end{aligned}
\end{align}
\hrule
\vspace{-12pt}
\end{figure*}

\subsection{Illustrations}

We now give instances of FICP to demonstrate the capability of above formulation to obtain optimal FIC (scalar or vector, linear or nonlinear) over the given alphabet.
\begin{example}
\label{ex_fldsz} Consider the FICP given in Table\,\ref{tab_fldsz} \cite{MinTx}.Here $n=1,\;K=4,\;N=6$. 
\begin{table}[h]
\setlength\extrarowheight{1pt}
\centering
\begin{tabular}{|c|c|c|}
\hline
\textbf{Client} & \textbf{Has-set} & \textbf{Want-set} \\
\hline
\textbf{$R_1$} & $\{x_1\;x_2\}$ & $\{x_3\;x_4\}$\\
\hline
\textbf{$R_2$} & $\{x_1\;x_3\}$ & $\{x_2\;x_4\}$\\
\hline 
\textbf{$R_3$} & $\{x_1\;x_4\}$ & $\{x_2\;x_3\}$\\
\hline
\textbf{$R_4$} & $\{x_2\;x_3\}$ & $\{x_1\;x_4\}$\\
\hline
\textbf{$R_5$} & $\{x_2\;x_4\}$ & $\{x_1\;x_3\}$\\
\hline
\textbf{$R_6$} & $\{x_3\;x_4\}$ & $\{x_1\;x_2\}$\\
\hline 
\end{tabular}
\vspace{5pt}
\caption{}
\label{tab_fldsz}
\vspace{-20pt}
\end{table}
The GELs are as follows:
\begin{align*}
\mathcal{M}(x_1,x_2,x_3,x_4)\neq \mathcal{M}(x_1,x_2,x_3',x_4'),\;\mathrm{if}\:(x_3,x_4)\neq (x_3',x_4')\\
\mathcal{M}(x_1,x_2,x_3,x_4)\neq \mathcal{M}(x_1,x_2',x_3,x_4'),\;\mathrm{if}\:(x_2,x_4)\neq (x_2',x_4')\\
\mathcal{M}(x_1,x_2,x_3,x_4)\neq \mathcal{M}(x_1,x_2',x_3',x_4),\;\mathrm{if}\:(x_2,x_3)\neq (x_2',x_3')\\
\mathcal{M}(x_1,x_2,x_3,x_4)\neq \mathcal{M}(x_1',x_2,x_3,x_4'),\;\mathrm{if}\:(x_1,x_4)\neq (x_1',x_4')\\
\mathcal{M}(x_1,x_2,x_3,x_4)\neq \mathcal{M}(x_1',x_2,x_3',x_4),\;\mathrm{if}\:(x_1,x_3)\neq (x_1',x_3')\\
\mathcal{M}(x_1,x_2,x_3,x_4)\neq \mathcal{M}(x_1',x_2',x_3,x_4),\;\mathrm{if}\:(x_1,x_2)\neq (x_1',x_2')
\end{align*}

Using Algorithm 1, we found that over $\mathbb{F}_2$, $L=L_{opt}=3$ while $L=L_{opt}=\mu(\mathcal{F})=2$ for both $\mathbb{F}_3$ and $\mathbb{F}_2^2$. This shows dependency of length of FIC on alphabet size and block length as asserted in \cite{ICRel,MinTx}. 

For $\mathbb{F}_2$, two maps are given without and in parentheses below. The former and the latter correspond to transmitting $(x_1,\, x_2+x_3,\, x_2+x_4)$ (linear) and $(x_1+x_4+ Maj(x_2,x_3,x_4),\, x_2+x_3,\, x_2+x_4)$ (non-linear) respectively. Thus, different assignments of codewords to color classes lead to different, possibly non-linear, codes.

\begin{small}
\begin{equation}
\label{gf2}
\begin{aligned}
\{0,7\}&\rightarrow 000\;(000) \qquad \qquad \{8,15\}\rightarrow 100\;(100)\\
\{1,6\}&\rightarrow 001\;(101) \qquad \qquad \{9,14\}\rightarrow 101\;(001)\\
\{2,5\}&\rightarrow 010\;(010) \qquad \qquad \{10,13\}\rightarrow 110\;(110)\\
\{3,4\}&\rightarrow 011\;(011) \qquad \qquad \{11,12\}\rightarrow 111\;(111)
\end{aligned}
\end{equation} 
\end{small}
For $\mathbb{F}_3$, a map is as follows:
\begin{small}
\begin{equation}
\label{gf3}
\begin{aligned}
\{0,16,23,35,39,46,58,65,78 \}&\rightarrow 00 \\
\{1,17,21,33,40,47,59,63,79 \}&\rightarrow 01 \\ 
\{2,15,22,34,41,45,57,64,80 \}&\rightarrow 02 \\
\{3,10,26,29,42,49,61,68,72 \}&\rightarrow 10 \\
\{4,11,24,27,43,50,62,66,73 \}&\rightarrow 11 \\
\{5,9,25,28,44,48,60,67,74  \}&\rightarrow 12 \\
\{6,13,20,32,36,52,55,71,75 \}&\rightarrow 20 \\
\{7,14,18,30,37,53,56,69,76 \}&\rightarrow 21 \\
\{8,12,19,31,38,51,54,70,77 \}&\rightarrow 22 \\
\end{aligned}
\end{equation} 
\end{small}
This corresponds to transmitting $(x_1+x_2+x_3,\,x_1+2x_2+x_4)$. For $\mathbb{F}_2^2$, a map is given below and corresponds to transmitting $(x_1^1+x_2^2+x_3^1,\, x_1^2+x_2^1+x_4^1,\, x_1^2+x_3^1+x_4^2,\, x_2^2+x_3^2+x_4^1)$.
{\fontsize{6}{7.2}\selectfont
\begin{align*}
\{0,29,38,59,71,90,97,124,137,148,175,178,206,211,232,245 \}&\rightarrow  0000\\
\{4,25,34,63,67,94,101,120,141,144,171,182,202,215,236,241 \}&\rightarrow  0001\\ 
\{1,28,39,58,70,91,96,125,136,149,174,179,207,210,233,244 \}&\rightarrow  0010\\
\{5,24,35,62,66,95,100,121,140,145,170,183,203,214,237,240 \}&\rightarrow  0011\\
\{6,27,32,61,65,92,103,122,143,146,169,180,200,213,238,243 \}&\rightarrow  0100\\
\{2,31,36,57,69,88,99,126,139,150,173,176,204,209,234,247 \}&\rightarrow  0101\\
\{7,26,33,60,64,93,102,123,142,147,168,181,201,212,239,242 \}&\rightarrow  0110\\
\{3,30,37,56,68,89,98,127,138,151,172,177,205,208,235,246 \}&\rightarrow  0111\\
\{9,20,47,50,78,83,104,117,128,157,166,187,199,218,225,252 \}&\rightarrow  1000\\
\{13,16,43,54,74,87,108,113,132,153,162,191,195,222,229,248 \}&\rightarrow  1001\\
\{8,21,46,51,79,82,105,116,129,156,167,186,198,219,224,253 \}&\rightarrow  1010\\
\{12,17,42,55,75,86,109,112,133,152,163,190,194,223,228,249 \}&\rightarrow  1011\\
\{15,18,41,52,72,85,110,115,134,155,160,189,193,220,231,250 \}&\rightarrow  1100\\
\{11,22,45,48,76,81,106,119,130,159,164,185,197,216,227,254 \}&\rightarrow  1101\\
\{14,19,40,53,73,84,111,114,135,154,161,188,192,221,230,251 \}&\rightarrow  1110\\
\{10,23,44,49,77,80,107,118,131,158,165,184,196,217,226,255 \}&\rightarrow  1111
\end{align*}}
For vector coding, the message vector is $(x_1^1,x_1^2,\ldots,x_4^1, x_4^2),x_i^j\in \mathbb{F}_2$. The vertices are labeled using the decimal equivalent of the message vector, \emph{e.g.}, $13=(1,1,0,1)$ in $\mathbb{F}_2$, $22=(0,2,1,1)$ in $\mathbb{F}_3$ and $198=(1,1,0,0,0,1,1,0)$ in $\mathbb{F}_2^2$. \hfill $\square$

\noindent
\textit{Remark:} We point out that, contrary to the authors' assertion, the ICP considered in \cite[Lemma 6]{MinTx} indeed has a scalar linear solution over $\mathbb{F}_2$, given by the following set of transmissions: $(x_1+x_2+x_3+x_5+x_7,\;x_4,\;x_5+x_6)$. 
\end{example}

%
%

\begin{example}
\label{ex_has} Consider the FICP descibed in Table\,\ref{tab_has}. Here, $x=(x_1,x_2,x_3,x_4)$.
\begin{table}[h]
\setlength\extrarowheight{1pt}
\centering
\begin{tabular}{|c|c|c|}
\hline
\textbf{Client} & \textbf{Has-set} & \textbf{Want-set} \\
\hline
\textbf{$R_1$} & $\{x_1\}$ & $\{x_2,x_3,x_4\}$\\
\hline
\textbf{$R_2$} & $\{x_2\}$ & $\{x_1,x_3,x_4\}$\\
\hline
\textbf{$R_3$} & $\{x_3\}$ & $\{x_1,x_2,x_4\}$\\
\hline
\textbf{$R_4$} & $\{x_4\}$ & $\{x_1,x_2,x_3\}$\\
\hline
\textbf{$R_5$} & $\mathit{Maj}(x_1,x_2,x_3)$ & $\{x_1,x_2,x_3,x_4\}$\\
\hline
\end{tabular}
\vspace{5pt}
\caption{}
\label{tab_has}
\vspace{-25pt}
\end{table}
The GELs are given in (\ref{el_has}). Executing our algorithm, we found that $L_{opt}=\mu(\mathcal{F})=3$ transmission are sufficient to satisfy all the demands. An encoding map is given below and corresponds to 

\vspace*{-10pt}
\begin{small}
\begin{align*}
\{0,15\}&\rightarrow 000 \qquad \qquad \{7,8\}\rightarrow 100\\
\{2,13\}&\rightarrow 001 \qquad \qquad \{5,10\}\rightarrow 101\\
\{4,11\}&\rightarrow 010 \qquad \qquad \{3,12\}\rightarrow 110\\
\{6,9\}&\rightarrow 011 \qquad \qquad \{1,14\}\rightarrow 111
\end{align*}
\end{small}
transmitting $(x_1+x_4,x_2+x_4,x_3+x_4)$. \hfill $\square$
\end{example}

\begin{example}
\label{ex_want} For the FICP given in Table\,\ref{tab_want}. Here $x=(x_1,x_2,x_3,x_4,x_5),\:x_i\in \mathbb{F}_2$. 
\begin{table}[h]
\setlength\extrarowheight{1pt}
\centering
\begin{tabular}{|c|c|c|}
\hline
\textbf{Client} & \textbf{Has-set} & \textbf{Want-set} \\
\hline
\textbf{$R_1$} & $\{x_1\;x_2\;x_3\}$ & $\{x_4,x_5\}$\\
\hline
\textbf{$R_2$} & $\{x_4\;x_5\}$ & $W_2(x)$\\
\hline
\end{tabular}
\vspace{5pt}
\caption{}
\label{tab_want}
\vspace{-15pt}
\end{table}
We consider 3 cases:\\
\textit{Case 1:} $W_2(x)=\mathit{Maj}(x_1,x_2,x_3)$\\ 
\textit{Case 2:} $W_2(x)=x_1+x_2+x_3$\\
\textit{Case 3:} $W_2(x)=(x_1,x_2,x_3) $

The GELs are given in (\ref{el_want}). The FIC size output by our method for the above cases are 4, 4 and 8 respectively, all of which are perfect. A map for Case 1 is

\vspace*{-10pt}
\begin{small}
\begin{align*}
\{0,4,8,13,16,21,25,29\}&\rightarrow 00 \; \{2,6,10,15,18,23,27,31\}\rightarrow 10\\
\{1,5,9,12,17,20,24,28\}&\rightarrow 01 \; \{3,7,11,14,19,22,26,30\}\rightarrow 11,
\end{align*}
\end{small}
\noindent
\hspace{-5pt}and corresponds to transmitting $(x_5+Maj(x_1,x_2,x_3),x_4+ x_5+ Maj(x_1,x_2,x_3))$, for Case 2 is 

\vspace*{-10pt}
\begin{small}
\begin{align*}
\{0,5,9,12,17,20,24,29\}&\rightarrow 00 \; \{2,7,11,14,19,22,26,31\}\rightarrow 01\\
\{1,4,8,13,16,21,25,28\}&\rightarrow 10 \; \{3,6,10,15,18,23,27,30\}\rightarrow 11
\end{align*}
\end{small}
\noindent
\hspace{-4pt}and corresponds to transmitting $(x_1+x_2+x_3+x_5,x_4)$, and for Case 3 is 

\vspace*{-10pt}
\begin{small}
\begin{align*}
\{0,9,18,27\}&\rightarrow 000 \qquad \{2,11,16,25\}\rightarrow 100\\
\{4,13,22,31\}&\rightarrow 001 \qquad \{6,15,20,29\}\rightarrow 101\\
\{1,8,19,26\}&\rightarrow 010 \qquad \{3,10,17,24\}\rightarrow 110\\
\{5,12,23,30\}&\rightarrow 011 \qquad  \{7,14,21,28\}\rightarrow 111
\end{align*}
\end{small}
\hspace{-4pt}and corresponds to transmitting $(x_1+x_4,+x_2+x_5,x_3)$.
\hfill $\square$
\end{example}

\section{Error-Correcting and Linear Functional Index Codes}
If the broadcast channel introduces noise, erroneous symbols may be received at the receivers. Let $\mathsf{wt}(\cdot)$ denote the Hamming weight of a vector. The results of this section generalize those given for error-correcting FSCSIP in \cite[Section~IV]{Anin} and for conventional ICP in \cite[Section~III and V]{Dau}.

\textit{Definition 5:} A $\delta$ error-correcting functional index code ($\delta$-FIC) for a given $\mathcal{F(X,R)}$ comprises of:\\
1. An encoding map, $\mathcal{M}:\mathbb{F}_q^{nK}\longrightarrow \mathcal{B},\,\mathcal{B}\subseteq \mathbb{F}_q^L$\\
2. Decoding functions, $\mathcal{D}_i:\mathcal{B} \times \mathbb{F}_q^{n|H_i|}\longrightarrow \mathbb{F}_q^{n|W_i|}$, such that $\forall i\in [N]$, $\forall x\in \mathbb{F}_q^{nK}$ and $\forall \epsilon\in \mathbb{F}_q^{L}$ such that $\mathsf{wt}(\epsilon)\leqslant \delta$, we have $\mathcal{D}(\mathcal{M}(x)+\epsilon,H_i(x))=W_i(x)$.

The following theorem states a necessary and sufficient condition for an encoding map to be a $\delta$-FIC for a given problem.

\begin{theorem} 
An encoding map $\mathcal{M}$ is a $\delta$-FIC for $\mathcal{F(X,R)}$ iff $\mathsf{wt}(\mathcal{M}(x)+\mathcal{M}(x'))\geqslant2\delta+1$, $\forall x,x'\in \mathbb{F}_q^{nK}$ such that $x$ and $x'$ are confusable.\\
\emph{Proof:} The Hamming ball of radius $\delta$ around message vector $x$, $B_H(x,\delta)\triangleq\{y\in\mathbb{F}_q^L\,\colon\,y=\mathcal{M}(x)+\epsilon,\epsilon\in \mathbb{F}_q^{L}, \mathsf{wt}(\epsilon)\leqslant \delta\}$ is the set of vectors obtained by introducing errors in at most $\delta$ coordinates in $x$. The correct decoding of $W_i(x), \forall i\in [N]$ is possible iff $B_H(x,\delta)\cap B_H(x',\delta)=\emptyset$ for every confusable pair $x,x'\in \mathbb{F}_q^{nK}$. If for all such pairs, $\mathcal{M}$ is a $\delta$-FIC, then we have that $\forall \epsilon, \epsilon'\in\mathbb{F}_q^L,\mathsf{wt}(\epsilon)\leqslant \delta$ and $\mathsf{wt}(\epsilon')\leqslant \delta $,

\begin{equation*} 
\mathcal{M}(x)+\epsilon\neq \mathcal{M}(x')+\epsilon'
\end{equation*}
or, 
\begin{equation*} 
\mathcal{M}(x)+\mathcal{M}(x')=\epsilon+\epsilon'
\end{equation*}

\noindent Since, $\{\epsilon+\epsilon'\colon \mathsf{wt}(\epsilon)\leqslant \delta,\mathsf{wt}(\epsilon')\leqslant \delta\}=\{\epsilon''\colon \mathsf{wt}(\epsilon'')\leqslant 2\delta \}$, we have, 
\begin{equation} \label{eq1}
\mathcal{M}(x)+\mathcal{M}(x')\neq\epsilon'',  \mathsf{wt}(\epsilon'')\leqslant 2\delta
\end{equation}
or, $\mathsf{wt}(\mathcal{M}(x)+\mathcal{M}(x'))\geqslant2\delta+1$. \hfill $\blacksquare$
\end{theorem}

The intuition behind this is that if the Hamming distance between the codewords of two confusable message vectors is $2\delta +1$, then at most $\delta$ errors can be corrected and the original codeword recovered. If the optimum code size for an FICP is $c$, then any classical error-correcting code with code size $c$ and minimum distance $2\delta +1$ can be used as a $\delta$-FIC for that FICP. If the error-correcting code used is optimal, \textit{i.e.}, has minimum block length given the code size and minimum distance, then the resulting $\delta$-FIC will also be optimal (minimum length FIC providing $\delta$ error-correction capability). Finding a minimum block length error-correcting code with a specified code size and minimum distance is NP-hard.

\begin{corollary} The FIC possesses no error-correcting capability when $\delta=0$, \emph{i.e.}, $\mathsf{wt}(\mathcal{M}(x)+\mathcal{M}(x'))\geqslant 1$ or $\mathcal{M}(x)\neq \mathcal{M}(x')$, for all confusable pairs $(x,x')$.
\end{corollary}
\noindent This is a restatement of Propositions 1 and 2.

\begin{corollary} A matrix $M$ is a $\delta$-FIC for $\mathcal{F(X,R)}$ iff $\mathsf{wt}((x+x')M)\geqslant2\delta+1$, $\forall x,x'\in \mathbb{F}_q^{nK}$ such that $x$ and $x'$ are confusable.\end{corollary}

\begin{corollary} For a linear FICP, a matrix $M$ is a $\delta$-FIC iff $\mathsf{wt}(xM)\geqslant2\delta+1$, $\forall x\in \mathbb{F}_q^{nK}$ such that $ x \in \cup _{i\in [N]}\{ \mathsf{Null}(M_{H_i}) \cap \{\mathbb{F}_q^{nK}\backslash \mathsf{Null}(M_{W_i})\}\}$.\\
\emph{Proof:} For the $i^{th}$ receiver, from Theorem~4, it follows that $\mathsf{wt}((x+x')M)\geqslant2\delta+1$, $\forall x,x'\in \mathbb{F}_q^{nK}$ such that $xM_{H_i}=x'M_{H_i}$ and $xM_{W_i}\neq x'M_{W_i}$, or, $(x+x')M_{H_i}=0$ and $(x+x')M_{W_i}\neq 0$. Substituting $x''$ for $(x+x')$, we have $\mathsf{wt}(x''M)\geqslant2\delta+1$, $\forall x''\in \mathbb{F}_q^{nK}$ such that $ x''M_{H_i}=0$ and $x''M_{W_i}\neq 0$. The result follows since this is true for all such $x\in \cup _{i\in [N]}\{ \mathsf{Null}(M_{H_i}) \cap \{\mathbb{F}_q^{nK}\backslash \mathsf{Null}(M_{W_i})\}\}$. \hfill $\blacksquare$
\end{corollary}

Note that the set $\cup _{i\in [N]}\{ \mathsf{Null}(M_{H_i}) \cap \{\mathbb{F}_q^{nK}\backslash \mathsf{Null}(M_{W_i})\}\}$ is the connection set of the confusion graph of a linear FICP (cf. Lemmas 1 and 2) and Corollary~4 states that if any matrix that maps message vectors in the connection set of the confusion graph to codewords of weight at least $2\delta+1$, then it represents a $\delta$-FIC.

\begin{theorem} The length, $L_{\delta}$, of a $\delta$-FIC is at least $L_{opt}+2\delta$, \emph{i.e.}, $L_{\delta}\geqslant L_{opt}+2\delta$.\\
\emph{Proof:} Let $c$ be as defined in Algorithm 1, \emph{i.e.}, the number distinct codewords required. Then $L_{opt}=\lceil \log_qc \rceil$, and the Hamming distance between any pair of codewords will be $1$. Let there be $c$ vectors of length $L_{\delta}$ over $\mathbb{F}_q$ such that the Hamming distance between any pair of vectors is at least $2\delta+1$. Puncturing all the vectors at arbitrary (but fixed) $2\delta$ coordinates, we will still have $c$ distinct vectors. Since minimum length of $c$ distinct vectors is $L_{opt}$, we have $L_{\delta}\geqslant L_{opt}+2\delta$. \hfill $\blacksquare$ \end{theorem} 

This is the Singleton bound for error-correcting FICs. Thus, concatenating an optimal $0$-FIC with an MDS code with minimum distance $2\delta +1$ will give an optimal $\delta$-FIC.

\begin{example}
\label{ex_ecc} Consider the FICP given in Table \ref{tab_ecc}. The FIC size output by Algorithm 1 is 2 (perfect) and transmitting $t=(x_1+x_4)(x_2+x_3)$ satisfies both the receivers. With this FIC, a $[2\delta +1,\,1,\,2\delta +1]$ repetition code, which is an MDS code, can be used as an outer code and the resultant code will be a $\delta$-FIC.
\begin{table}[h]
\setlength\extrarowheight{1pt}
\centering
\begin{tabular}{|c|c|c|}
\hline
\textbf{Client} & \textbf{Has-set} & \textbf{Want-set} \\
\hline
\textbf{$R_1$} & $Maj(x_2,x_3,x_4)$ & $\{Maj(x_1,x_2,x_3)$,\\
 & & $\,x_1^c(x_2+x_3)x_4\}$\\
\hline
\textbf{$R_2$} & $Maj(x_1+x_3,x_2,x_4)$ & $(x_1+x_2)x_3+x_1x_4$\\
\hline
\end{tabular}
\vspace{5pt}
\caption{}
\label{tab_ecc}
\vspace{-10pt}
\end{table}

An optimal linear FIC of length $2$ is $t_1=x_1+x_4,\,t_2=x_2+x_3$ and the transmissions of an optimal $1$-FIC are $(t_1,t_1, t_2,t_2, t_1+t_2)$. The matrices $M_0$ and $M_1$ for $0$- and $1$-FIC respectively are given below.
\[ M_0= \begin{bmatrix}
1 & 0\\
0 & 1\\
0 & 1\\
1 & 0\\
\end{bmatrix}
\qquad \quad
M_1= \begin{bmatrix}
1 & 1 & 0 & 0 & 1\\
0 & 0 & 1 & 1 & 1\\
0 & 0 & 1 & 1 & 1\\
1 & 1 & 0 & 0 & 1\\
\end{bmatrix} 
\]
As stated before, the minimum distance of these codes are 1 and 3 respectively.
\hfill $\square$
\end{example}

\begin{example}
For the FICP of Table~\ref{tab_fldsz}, for coding over $\mathbb{F}_2$, a $[6,3,3]$ code (obtained by shortening the $[7,4,3]$ Hamming code) can be used to obtain an optimal $1$-FIC ($M_1$); the matrix representations of the linear code given in \eqref{gf2} (without parentheses) are 
\[ M_0 = \begin{bmatrix}
1 & 0 & 0\\ 
0 & 1 & 1\\
0 & 1 & 0\\ 
0 & 0 & 1\\
\end{bmatrix}
\qquad
G = \begin{bmatrix}
1 & 0 & 0 & 1 & 1 & 0\\
0 & 1 & 0 & 1 & 0 & 1\\
0 & 0 & 1 & 0 & 1 & 1\\
\end{bmatrix},
\]
where $M_1 = M_0\,G$. This code does not satisfy the Singleton bound. 

For coding over $\mathbb{F}_3$, we use the codewords of a $[4,2,3]$ MDS code over $\mathbb{F}_3$ ($G$) instead of those mentioned in \eqref{gf3} to obtain an optimal $1$-FIC ($M_1$); the codewords are $\{0000,\,0111,\,0222,\,1012,\,1120,\,1201,\,2021,\,2102,\,2210\}$; the matrix representations are
\[ M_0 = \begin{bmatrix}
1 & 1\\ 
1 & 2\\
1 & 0\\ 
0 & 1\\
\end{bmatrix}
\qquad
G = \begin{bmatrix}
1 & 0 & 1 & 2\\
0 & 1 & 1 & 1\\
\end{bmatrix},
\]
where $M_1 = M_0\,G$. This code satisfies the Singleton bound. 

For coding over $\mathbb{F}_2^2$, the $[7,4,3]$ Hamming code ($G$) can be concatenated to the $0$-FIC ($M_0$) to get a $1$-FIC ($M_1$); the matrix representations are
\[ M_0 = \begin{bmatrix}
1 & 0 & 0 & 0\\ 
0 & 1 & 1 & 0\\
0 & 1 & 0 & 0\\ 
1 & 0 & 0 & 1\\
1 & 0 & 1 & 0\\ 
0 & 0 & 0 & 1\\
0 & 1 & 0 & 1\\ 
0 & 0 & 1 & 0\\
\end{bmatrix}
\qquad
G = \begin{bmatrix}
1 & 0 & 0 & 0 & 1 & 1 & 0\\
0 & 1 & 0 & 0 & 1 & 0 & 1\\
0 & 0 & 1 & 0 & 0 & 1 & 1\\
0 & 0 & 0 & 1 & 1 & 1 & 1\\
\end{bmatrix},
\]
where $M_1 = M_0\,G$. This code also does not satisfy the Singleton bound. \hfill $\square$
\end{example}

\section{Discussion}
A novel extension of the ICP was proposed wherein client nodes can hold as side information and demand functions of messages generated by the transmitter. An FIC should be such that, given the coded transmissions by source and the side information, each client should be able to resolve what the value of its demanded information is. The restriction posed by each client on the FIC was formulated via GELs. Based on these GELs, a graph was constructed and it was shown that a vertex coloring of this graph gives a valid FIC. Some properties of the confusion graph and bounds on the optimal code size were subsequently obtained. Illustrations were provided to attest that the devised method to obtain a FIC provides an optimal solution over the given alphabet. Transmission over noisy broadcast channel was then studied and a necessary and sufficient condition for an FIC to be $\delta$ error-correcting and lower bound on length of a $\delta$-FIC were subsequently obtained.

Topics of further study include method of obtaining linear FICs (not necessarily optimal), identifying FICPs with efficiently colorable confusion graphs and studying and exploiting the structure of confusion graphs to facilitate coloring and applying heuristic and approximation algorithms for the same.  


\end{document}